\documentclass[12pt]{iopart}
\usepackage{bm}
\usepackage{harvard}
\usepackage{amssymb}
\usepackage
{hyperref}

\begin{document}
\title[Alpha variation problem and $q$-factor definition]
{Alpha variation problem and $q$-factor definition}

\author{K V Koshelev}

\address{ Petersburg Nuclear Physics Institute,
Gatchina 188300, Russia}

\ead{koshelev@landau.phys.spbu.ru}

\date{\today }

\begin{abstract}
Different $q$-factor definitions are considered. The formula
connecting different $q$-factors is given. Also it is pointed the
way to find all the $q$-factors from experimental data.
\end{abstract}




\section{Introduction}
\label{intro}
At present time the problem of $\alpha=\frac{e^2}{\hbar c}$
(fine-structure constant) variation is of great interest for many
scientists \cite{Uzan}. To extract information about $\alpha$
variation from the atomic experimental data one needs so called
sensitivity coefficients or $q$-factors for the transition
frequencies. In general the $q$-factor is just the derivative of
the transition frequency over one of the fundamental constants
such as velocity of the light $c$ (the usual choice), the Planck
constant $\hbar$ or elementary charge $e$. The calculation of the
$q$-factors is a severe task in general case. It is usually done
by the numerical differentiation of the energies \cite{GL74}.
Therefore it is very important to find method to estimate
$q$-factors from the experimental data (without any atomic
structure calculations). The main idea of the report is to apply
for the $q$-factor calculation the Hell-Mann-Feynman theorem.

\section{Different $q$-factors definitions and connections between them}
\label{basic}
Let's take the relativistic hamiltonian of the atomic system
in the form
\begin{equation}\label{hamiltonian0}
H=h_1+h_2+h_3
\end{equation}
where
\begin{equation}\label{hamiltonian1}
h_1=\sum c (\overrightarrow \alpha \cdot \overrightarrow p)
\end{equation}
\begin{equation}\label{hamiltonian2}
h_2=\sum \left[\frac{e^2}{r_{12}}-\frac{e^2}{2}
 \left(\frac{(\overrightarrow\alpha_1\cdot\overrightarrow\alpha_2)}
{r_{12}}+\frac{(\overrightarrow\alpha_1\cdot\overrightarrow
r_{12})(\overrightarrow\alpha_2\cdot\overrightarrow r_{12})}
{r_{12}^3}\right)\right]-\sum \frac{e^2 Z}{r}
\end{equation}
\begin{equation}\label{hamiltonian3}
h_3=\sum m c^2 \beta
\end{equation}
The term $h_1$ corresponds to the kinetic energy of the system of
the electrons, the term $h_2$ presents the Coulomb interaction of
the electrons with atomic nucleus and interelectron interaction
(both Coulomb and Breit) between electrons and finally the $h_3$
term is the rest mass term. Applying the Hell-Mann-Feynman theorem
to the diagonal matrix element of the hamiltonian
(\ref{hamiltonian0}) one can easy receive
\begin{equation}\label{q-factors1}
c\frac{\partial\langle H\rangle}{\partial c}=\langle
h_1\rangle+2\langle h_3\rangle
\end{equation}
\begin{equation}\label{q-factors2}
\hbar\frac{\partial\langle H\rangle}{\partial \hbar}=\langle
h_1\rangle
\end{equation}
\begin{equation}\label{q-factors3}
e^2\frac{\partial\langle H\rangle}{\partial e^2}=\langle
h_2\rangle
\end{equation}
Let's introduce the definitions $q_c=c\frac{\partial\langle
H\rangle}{\partial c}$, $q_\hbar=\hbar\frac{\partial\langle
H\rangle}{\partial \hbar}$ and $q_e=e^2\frac{\partial\langle
H\rangle}{\partial e^2}$ for the velocity $q_c$-factor, Planck
constant $q_\hbar$-factor and elementary charge $q_e$-factor
respectively. As a consequence of the formulae (\ref{q-factors1},
\ref{q-factors2}, \ref{q-factors3}) we have the relation of the
$q$-factors
\begin{equation}\label{q-factors4}
\langle H\rangle=\frac{1}{2}(q_c+q_\hbar)+q_e
\end{equation}
The virial theorem yields
\begin{equation}\label{virial}
\langle h_1\rangle=-\langle h_2\rangle=q_\hbar=-q_e
\end{equation}
As a consequence of the formulae (\ref{q-factors1},
\ref{q-factors4}, \ref{virial}) we have
\begin{equation}\label{c}
\langle H\rangle=\langle h_3\rangle
\end{equation}
Taking into consideration formulae (\ref{q-factors1}, \ref{c}) and
$q$-factors definitions one can easy receive for the $q_c$
\begin{equation}\label{he}
q_c=q_\hbar+2E
\end{equation}
where $E=\langle H\rangle$ is  the energy of the reference state.
It's easy to see that formulae for the $q$-factors given above are
valid not only for one particular energy level $q$-factors but
also for the difference of the $q$-factors (transition
$q$-factor). In the latter case one needs to change the term $2E$
in the formula (\ref{he}) to $2\omega$ (doubled transition
frequency). So in the following we'll not make the difference
between one particular energy level $q$-factor (energy of the
state) and transition $q$-factor (transition frequency).

\section{Approximate formulae for the $q$-factors}
It's clear that to find all three $q$-factors it is enough to
receive just one of them. Omitting the operators corresponding to
the  interelectron interaction in the formula (\ref{hamiltonian2})
we have the following approximation for the $q_e$-factor
\begin{equation}\label{approximation}
q_e\simeq\left\langle-\sum \frac{e^2 Z}{r}\right\rangle
\end{equation}
The matrix element in formula (\ref{approximation}) looks like the
first order perturbation theory correction to the energy of the
reference state due to the Coulomb interaction of the electrons
with nucleus (with factor $Z$ omitted). This correction may be
extracted from the experimental data. For example one can compare
the spectral lines data for the ion of the interest and one
elementary charge higher nucleus ion data (with the same number of
electrons). Therefore one can write the next approximation
\begin{equation}\label{approximation1}
q_e\simeq Z\left(E_{Z+1}-E_Z\right)
\end{equation}
Where the $E_{Z+1}$ and $E_Z$ are the energy of the reference
state perturbed by Coulomb potential (in the ion with the charge
of the nucleus amounts to $Z+1$) and reference state energy (in
the ion with the charge of the nucleus amounts to $Z$)
respectively. Combining together the formulae
(\ref{virial}),(\ref{he}) and (\ref{approximation1}) one can easy
get the final approximate formula for the $q_c$-factor
\begin{equation}\label{approximation2}
q_c\simeq (Z+2)E_{Z}-Z E_{Z+1}
\end{equation}
Several approximations were made to receive the formulae
(\ref{approximation1}) and (\ref{approximation2}). The possibility
and quality of these approximations is the question in each
particular atomic system case.

\section{Conclusions}
All the possible $q$-factors are connected each other
(corresponding formula is given). Application of the formulae
(\ref{approximation1}), (\ref{approximation2}) gives the
possibility to find all the $q$-factors from experimental data.

\end{document}